\documentclass[amsmath,amssymb,12pt,superscriptaddress,nofootinbib]{revtex4-1}

\usepackage{graphicx}
\usepackage{dcolumn}
\usepackage{bm}
\usepackage{color}
\usepackage{enumitem}
  \usepackage{tabularx}
   \newcolumntype{C}{>{\centering\arraybackslash}X}
   \newcolumntype{L}{>{\raggedright\arraybackslash}X}
   \newcolumntype{R}{>{\raggedleft\arraybackslash}X}
\setlistdepth{10}

\newcommand{\dd}{\mathrm{d}}
\newcommand{\rg}{r_{\rm g}}
\newcommand{\rmc}{{\rm h}}
\newcommand{\del}{\partial}

\definecolor{DarkBlue}{rgb}{0,0,0.7} 

\definecolor{DarkRed}{rgb}{0.65,0,0}

\begin{document}
\baselineskip5.5mm


{\baselineskip0pt
\small
\leftline{\baselineskip16pt\sl\vbox to0pt{
               \hbox{\it Division of Particle and Astrophysical Science, Nagoya University}
               \hbox{\it Department of Mathematics and Physics, Osaka City University}
                             \vss}}
\rightline{\baselineskip16pt\rm\vbox to20pt{
 		\hbox{OCU-PHYS 493}
        	\hbox{AP-GR 152}
	\hbox{NITEP 4}
\vspace{-1.5cm}
\vss}}
}

\author{Chul-Moon~Yoo}\email{yoo@gravity.phys.nagoya-u.ac.jp}
\affiliation{
Division of Particle and Astrophysical Science,
Graduate School of Science, Nagoya University, 
Nagoya 464-8602, Japan
\vspace{1.5mm}
}

\author{Ken-ichi~Nakao}\email{knakao@sci.osaka-cu.ac.jp}
\affiliation{
Department of Mathematics and Physics, Graduate School of Science, Osaka City University, 3-3-138 Sugimoto, Sumiyoshi, Osaka 558-8585, Japan
\vspace{1.5mm}
}

\vskip2cm
\title{Constant-mean-curvature Slicing of the Swiss-cheese Universe
}

\begin{abstract}
\vskip1cm 
A sequence of Constant-Mean-Curvature(CMC) slices in the Swiss-Cheese(SC) Universe is investigated. 
We focus on the CMC slices which smoothly connect to the homogeneous time slices in 
the Einstein-de~Sitter region in the SC universe. 
It is shown that the slices do not pass through the black hole region but white hole region. 
\end{abstract}

\maketitle

\section{Introduction}
\label{sec:intro}
Numerical simulations of spacetime dynamics in cosmological settings 
have been 
actively performed in recent years. 
One main motivation to simulate the cosmological nonlinear dynamics is 
to quantify the effect of the non-linear small scale inhomogeneity on the global 
expansion law of the universe\cite{RevModPhys.29.432,Clifton:2009jw,Clifton:2012qh,Bentivegna:2012ei,Yoo:2012jz,Bentivegna:2012ei,Bruneton:2012cg,Bruneton:2012ru,Bentivegna:2013xna,Yoo:2013yea,Bentivegna:2013jta,Clifton:2013jpa,Korzynski:2013tea,Clifton:2014lha,Yoo:2014boa,Ikeda:2015hqa,Bentivegna:2016fls}. 
Another significant motivation comes from primordial black holes\cite{1967SvA....10..602Z,Hawking:1971ei}. 
Spherically symmetric simulations of primordial black hole formation have 
been 
repeated in different settings\cite{1978SvA....22..129N,1980SvA....24..147N,Shibata:1999zs,Niemeyer:1999ak,Musco:2004ak,Polnarev:2006aa,Musco:2012au,Polnarev:2012bi,Nakama:2013ica,Nakama:2014fra}. 
Non-spherical simulation of gravitational collapse in an expanding background has been recently performed 
in Ref.~\cite{Yoo:2018pda}. 

When we analyze a spacetime dynamics with a numerical procedure, 
the dynamics is described as a sequence of time slices, 
that is, 
a foliation by a one-parameter family of spacelike hypersurfaces. 
Therefore, in order to understand the spacetime structure,  
the domain covered by the sequence of time slices 
should be correctly figured out. 
For this purpose, a sequence of time slices in a well-known analytic spacetime is often helpful. 
One of the useful time slice conditions is the so-called Constant-Mean-Curvature(CMC), 
which requires a uniform value of the trace of the extrinsic curvature 
of each time slice.   
A CMC slice is often taken as the initial hypersurface 
for numerical simulations because it simplifies the Hamiltonian and momentum constraint equations 
under certain assumptions. 
CMC slices in the Schwarzschild(Sch) spacetime may give a helpful insight 
to understand the intrinsic geometry and the embedding of 
the initial hypersurface for a dynamical simulation associated with black hole formation. 

There are several works on CMC and other slices for well-known 
spacetimes(see e.g., \cite{Estabrook:1973ue,Nakao:1990gw,Beig:2005ef,Nakao:2009dc,Dennison:2014eta,Dennison:2017mqd}). 
In this paper, we investigate CMC slices in the Swiss-Cheese(SC) universe\cite{Einstein:1945id,Einstein:1946zz}. 
The SC universe model is constructed by arbitrarily removing spherical regions from the Einstein-de Sitter (EdS) universe model in a non-overlapping manner, 
and filling each removed region with a region of the Sch spacetime whose center is occupied by a black or white hole. 
Thus, the SC universe is composed of the interior Sch region and exterior 
EdS region which 
are matched each other with 
Israel's junction condition\cite{Israel:1966rt}. 
In the EdS region, we consider the trivial CMC slice, that is, the homogeneous slice 
on which the value of the trace of the extrinsic curvature is given by $3H$ with $H$ being the Hubble constant. 
Therefore, what we investigate in this paper is just a sequence of CMC slices in the Sch spacetime. 
The difference from previous studies on CMC slices in the Sch spacetime is in the boundary condition 
on the boundary 
between the Sch and EdS regions. 
The previous studies on a foliation of a black hole spacetime with CMC slices 
are applicable to only totally spherically symmetric spacetimes.
By contrast, the CMC slices in the present study will be applicable to 
the situations in which black holes are randomly distributed in the expanding universe. 
The knowledge about the CMC slices in the SC universe may be helpful to 
get better insight into the geometry of the initial hypersurface. 

This paper is organized as follows. 
In Sec.~\ref{sec:matching}, we review the SC universe deriving the equation 
describing the boundary between the Sch and EdS regions. 
The ordinary differential equations for CMC slices with the Kruskal coordinates 
are derived in Sec.~\ref{sec:odes}, and results are shown in Sec.~\ref{sec:result}. 

We use the geometrized units in which both
the speed of light and Newton's gravitational constant are 
one. 
The Greek indices run from 0 to 3 and the Latin indices run from 1 to 3.

\section{Matching EdS and the Sch spacetime}
\label{sec:matching}

As mentioned, the SC universe model is constructed by removing spherical regions from the EdS universe model and filling each removed domain by 
a spherical domain of the Sch spacetime. 
We briefly review the SC universe model to fix the notation in this section 
deriving the 
motion of a boundary 
between the EdS and the Sch regions. 

\subsection{Boundary on the EdS side}
The line element of the EdS spacetime can be written as 
\begin{equation}
\dd s^2=-\dd\tau^2+a(\tau)^2\left(\dd\chi^2+\chi^2\dd\Omega^2\right), 
\end{equation}
where $a(\tau)$ is the scale factor, and $\dd\Omega^2=\dd\theta^2+\sin^2\theta \dd\phi^2$ is the round metric. 
If we set $a=a_\rmc$ at $\tau=\tau_\rmc$, 
we have 
\begin{equation}
a(\tau)=a_\rmc \left(\frac{\tau}{\tau_\rmc}\right)^{2/3}. 
\end{equation}

The one-parameter family of the timelike hypersurfaces with constant $\chi$ 
foliate the EdS spacetime. 
On each 
hypersurface of constant $\chi$, we use the intrinsic coordinates $\xi^i$ defined by 
\begin{equation}
\xi^i=(\tau,\theta,\phi). 
\end{equation}
Then, the induced metric $h_{ij}$ on a 
hypersurface of constant $\chi$ is given by 
\begin{equation}
h_{ij}\dd\xi^i\dd\xi^j=-\dd\tau^2+a(\tau)^2\chi^2\dd\Omega^2.  
\end{equation}

We remove a spherical region $\chi<\chi_{\rm b}$ in the EdS universe model and fill it with a region of the Sch spacetime. 
The boundary between the Sch and the EdS regions is a timelike hypersurface $\Sigma_0$ of  
\begin{equation}
\chi=\chi_{\rm b}. 
\label{eq:sphere}
\end{equation}
Defining $A(\tau)$ as 
\begin{equation}
A(\tau):=a(\tau)\chi_{\rm b}=A_\rmc\left(\frac{\tau}{\tau_\rmc}\right)^{2/3}, 
\label{eq:area}
\end{equation}
we obtain the following form of the induced metric $h_{ij}^{\rm b}$ 
on $\Sigma_0$: 
\begin{equation}
h_{ij}^{\rm b}\dd\xi^i\dd\xi^j=-\dd\tau^2+A(\tau)^2\dd\Omega^2, 
\end{equation}
where we have defined $A_\rmc$ by $A_\rmc=a_\rmc \chi_{\rm b}$. 

As is well known, for the Gaussian normal coordinate, 
the extrinsic curvature tensor $k_{ij}$ of $\Sigma_0$ 
is given by 
\begin{equation}
k_{ij}=\frac{1}{2}\left.\ell^\mu\del_\mu h_{ij}\right|_{\chi=\chi_{\rm b}}, 
\end{equation}
where 
$\ell^\mu$ is the normalized vector 
which is normal to $\Sigma_0$. 
Since we have 
\begin{equation}
\ell^\mu\del_\mu=\frac{1}{a}\del_\chi,  
\end{equation}
nonzero components of the extrinsic curvature 
are given by 
\begin{equation}
k_{\theta\theta}=\frac{k_{\phi\phi}}{\sin^2\theta}=A(\tau).  
\end{equation}

\subsection{Boundary on the Sch side}
\label{sec:bsch}
The metric of the Sch spacetime is given by 
\begin{equation}
\dd s^2=-f(r)\dd t^2+\frac{1}{f(r)}\dd r^2+r^2\dd \Omega^2, 
\label{eq:Schmet}
\end{equation}
where 
\begin{equation}
f(r)=1-\frac{2M}{r}. 
\end{equation}
The boundary between the EdS and the Sch regions, $\Sigma_0$, is described in the Sch side in the following manner: 
\begin{equation}
t=t(\tau)~,~r=r(\tau)~,~\theta=\theta~,~\phi=\phi. 
\end{equation}
The induced metric is given by 
\begin{equation}
h^{\rm b}_{ij}\dd\xi^i\dd\xi^j
=\left(-f(r)t'^2+\frac{r'^2}{f(r)}\right)\dd \tau^2+r^2\dd \Omega^2, 
\end{equation}
where $t'=\dd t/\dd \tau$ and $r'=\dd r/\dd\tau$. 
From the 1st Israel Junction condition (equivalence of $h_{ij}^{\rm b}$), 
we obtain 
\begin{eqnarray}
&&-ft'^2+\frac{r'^2}{f}=-1,
\label{eq:normetau}\\
&&r=A_\rmc \left(\frac{\tau}{\tau_\rmc}\right)^{2/3}. 
\label{eq:rsol}
\end{eqnarray}

Setting $x^\mu=(t,r,\theta,\phi)$ and 
$e^\mu_i:=\del x^\mu/\del\xi^i$, 
we have the expression for the extrinsic curvature $k_{ij}$ as 
\begin{equation}
k_{ij}=-\ell_\mu e^\nu_i\nabla_\nu e^\mu_j. 
\end{equation}
From this expression, we obtain 
\begin{equation}
k_{\theta\theta}=\frac{1}{2}\ell_rg^{rr}\partial_rg_{\theta\theta}. 
\end{equation}
Since 
\begin{equation}
\ell_\mu=(-r',t',0,0), 
\end{equation}
we find
\begin{equation}
k_{\theta\theta}=t'rf(r). 
\end{equation}

In this paper, we do not allow any singular shell source on the boundary. 
Therefore, from the 2nd Junction condition, 
$k_{ij}$ must 
have an identical value on each 
side. 
Then, comparing the values of $k_{\theta\theta}$, we obtain 
\begin{equation}
t'=\frac{1}{f(r)}
=\left(1-\frac{2M}{A_\rmc}\left(\frac{\tau_\rmc}{\tau}\right)^{2/3}\right)^{-1}.
\label{eq:dtsol}
\end{equation}
The other non-trivial component is $k_{\tau\tau}$: 
\begin{equation}
k_{\tau\tau}=-
\ell_\mu e^\nu_\tau\nabla_\nu e^\mu_\tau. \label{eq:ktautau}
\end{equation}
Equation \eqref{eq:normetau} is equivalent to $g_{\mu\nu}e_\tau^\mu e_\tau^\mu=-1$ and hence we have
$$
g_{\alpha\beta}e_\tau^\alpha e^\nu_\tau\nabla_\nu e^\beta_\tau=0.
$$
Furthermore, we can easily find $e^\nu_\tau\nabla_\nu e^\theta_\tau=0=e^\nu_\tau\nabla_\nu e^\phi_\tau$. 
Thus, Eq.~\eqref{eq:ktautau} and the 2nd Junction condition on $\tau$-$\tau$ component, $k_{\tau\tau}=0$, lead to
\begin{equation}
 e^\nu_\tau\nabla_\nu e^\mu_\tau=0. \label{eq:geodesic}
\end{equation}
This equation is just a geodesic equation.

The $r$ component of Eq.~\eqref{eq:geodesic} leads to
\begin{equation}
r''+\frac{1}{2}\del_rf=0,
\end{equation}
where we have used Eq.~\eqref{eq:normetau}.
Using Eq.~\eqref{eq:rsol}, we obtain 
\begin{equation}
\frac{2}{9}A_\rmc^3\tau_\rmc^{-2}=M. 
\end{equation}
This condition 
implies that the mass inside 
the sphere 
specified by Eq.~\eqref{eq:sphere} in EdS 
is equivalent to the mass of the Sch spacetime. 
Actually, we can confirm it as follows: 
\begin{equation}
M=\rho\times \frac{4}{3}\pi A^3=\frac{3}{8\pi}H^2\times \frac{4}{3}\pi A^3
=\frac{2}{9}A_\rmc^3\tau_\rmc^{-2}, \label{eq:mass}
\end{equation}
where $\rho$ and $H\equiv a'/a=A'/A$ 
are the energy density and the Hubble constant, respectively, and we have used the Friedmann equation 
$$
H^2=\frac{8}{3}\pi\rho. 
$$
From the $t$-component of the geodesic equation, 
we obtain
\begin{equation}
t''+\frac{1}{f^2}\del_rfr'=0. 
\end{equation}
We can check that this condition is automatically satisfied by 
\eqref{eq:dtsol}. 

Let us define $\tau_\rmc$ such that 
$r=2M$ at $\tau=\tau_\rmc$, so that the area radius $r$ of the boundary between the EdS and the Sch regions 
is larger than $2M$ for $\tau>\tau_\rmc$, whereas it is less than or equal to $2M$ for $\tau\leq \tau_\rmc$. 
Then, we obtain 
\begin{equation}
A_\rmc=2M=\frac{3}{2}\tau_\rmc, 
\end{equation}
where we have used Eq.~\eqref{eq:mass}. 
Finally, we obtain 
\begin{align}
r&=\frac{3}{2}\tau_\rmc\left(\frac{\tau}{\tau_\rmc}\right)^{2/3}, 
\label{eq:rsol2}
\\
t'&=\left[1-\left(\frac{\tau_\rmc}{\tau}\right)^{2/3}\right]^{-1}. 
\label{eq:dtdtau}
\end{align}
The second equation can be integrated as 
\begin{align}
t&=\tau+3\tau_\rmc\left(\frac{\tau}{\tau_\rmc}\right)^{1/3}
+\frac{3}{2}\tau_\rmc\ln\left[\frac{-1+\left(\frac{\tau}{\tau_\rmc}\right)^{1/3}}
{1+\left(\frac{\tau}{\tau_\rmc}\right)^{1/3}}\right]\cr
&=\tau+3\tau_\rmc\left(\frac{\tau}{\tau_\rmc}\right)^{1/3}
-3\tau_\rmc {\rm Arccoth}\left[\left(\frac{\tau}{\tau_\rmc}\right)^{1/3}\right], 
\label{eq:tsol}
\end{align}
where we have omitted the integration constant which 
can be chosen freely without loss of generality 
because of the time translational invariance. 
Eventually, we have only one parameter $\tau_\rmc$.

\subsection{Kruskal extension}
So far, we derived every equation based on 
the Schwarzschild coordinate system given in Eq.~\eqref{eq:Schmet}. 
Since the single patch of this coordinate system covers only either the outside of the horizon $r>2M$
or the inside of the horizon $r<2M$, 
we consider an analytic extension beyond the horizon $r=2M$. 
The Kruskal coordinates are well known as a coordinate system 
which covers the whole spacetime region. 
For later convenience, let us consider the 
Kruskal extension of the trajectory of $\Sigma_0$ 
in the Sch spacetime. 
Outside the horizon, 
the Kruskal coordinates and the coordinates $(t,r)$ 
are related as 
\begin{eqnarray}
T&=&\frac{1}{2}\left(\exp\left[\frac{t+r+\rg\ln\left[(r-\rg)/\rg\right]}
{2\rg}\right]
-\exp\left[\frac{-t+r+\rg\ln\left[(r-\rg)/\rg\right]}
{2\rg}\right]
\right), \\
R&=&\frac{1}{2}\left(\exp\left[\frac{t+r+\rg\ln\left[(r-\rg)/\rg\right]}
{2\rg}\right]
+\exp\left[\frac{-t+r+\rg\ln\left[(r-\rg)/\rg\right]}
{2\rg}\right]
\right). 
\end{eqnarray}
Substituting Eqs.~\eqref{eq:rsol2} and \eqref{eq:tsol} into 
these expressions, we get 
\begin{eqnarray}
T_{\rm b}(\tau)&=&\exp\left[\frac{1}{2}\left(\frac{\tau}{\tau_\rmc}\right)^{2/3}\right]
\left\{\left(\frac{\tau}{\tau_\rmc}\right)^{1/3}
\sinh\left[\left(\frac{\tau}{\tau_\rmc}\right)^{1/3}
+\frac{\tau}{3\tau_\rmc}\right]-\cosh\left[
\left(\frac{\tau}{\tau_\rmc}\right)^{1/3}+\frac{\tau}{3\tau_\rmc}\right]
\right\}, ~~~~~
\label{eq:Tb}
\\
R_{\rm b}(\tau)&=&\exp\left[\frac{1}{2}\left(\frac{\tau}{\tau_\rmc}\right)^{2/3}\right]
\left\{\left(\frac{\tau}{\tau_\rmc}\right)^{1/3}\cosh\left[
\left(\frac{\tau}{\tau_\rmc}\right)^{1/3}+\frac{\tau}{3\tau_\rmc}\right]
-\sinh\left[\left(\frac{\tau}{\tau_\rmc}\right)^{1/3}
+\frac{\tau}{3\tau_\rmc}\right]\right\}. 
\label{eq:Rb}
\end{eqnarray}
We note that the calculations in Sec.~\ref{sec:bsch} can be also applied to the region $r<2M$ and 
we obtain the same expression Eq.~\eqref{eq:tsol} for the trajectory. 
Therefore, the expressions \eqref{eq:Tb} and \eqref{eq:Rb} 
can be analytically extended 
to the region $\tau<\tau_\rmc$. 

As is clear from Eq.~\eqref{eq:area}, 
the area radius $A(\tau)$ vanishes at $\tau=0$ and diverges
for $\tau\rightarrow\infty$. 
Such a timelike trajectory is possible in Sch spacetime only for the case 
starting from the past singularity and going to the future timelike infinity. 
Therefore, the boundary trajetory 
necessarily passes through the white hole horizon(see Fig.~\ref{fig:Kruskal} in Sec.~\ref{sec:result}). 

\section{Differential equations for CMC slices}
\label{sec:odes}

We are interested in CMC slices, that is, spacelike hypersurfaces with constant $K$ 
in the SC universe, where $K$ is the trace of the extrinsic curvature of the spacelike hypersurface. 
In the EdS region, we choose 
homogeneous time slices. 
Then, we extend these time slices to the Sch region 
keeping $K=$const. 
For this purpose, we derive the 
differential equations for CMC slices 
in the Sch spacetime. 

We use the Kruskal coordinate system, in which 
the line element is written as 
\begin{equation}
\dd s^2=\frac{4\rg^3}{r(T,R)}\exp\left(-\frac{r(T,R)}{\rg}\right)\left(-\dd T^2+\dd R^2\right)
+r(T,R)^2\dd \Omega^2, 
\end{equation}
where $r(T,R)$ is defined by 
\begin{equation}
T^2-R^2=-\left(\frac{r-\rg}{\rg}\right)\exp\left(\frac{r}{\rg}\right). 
\label{eq:defr}
\end{equation}

We consider a spacelike hypersurface $\Sigma_1$ 
specified by the following 
parametric equation:
\begin{eqnarray}
T&=&f_T(v),\\
R&=&f_R(v). 
\end{eqnarray}
Covariant components of the vector normal to this surface is given by 
\begin{equation}
N_\mu=\pm\left(-\dot f_R~,~ \dot f_T~,~ 0~,~ 0\right), 
\end{equation}
where we have chosen $\pm$ such that $N^\mu$ 
is future directed, i.e., 
$+$ for $\dot f_R>0$ and 
$-$ for $\dot f_R<0$. 
The dot ``$\dot ~$" denotes $\dd/\dd v$. 
The contravariant 
components are given by 
\begin{equation}
N^\mu=\pm\frac{r}{4\rg^3}\exp\left(\frac{r}{\rg}\right)
\left(\dot f_R~,~ \dot f_T~,~0~,~0\right). 
\end{equation}
The norm is calculated as 
\begin{equation}
N_\mu N^\mu=\frac{r}{4\rg^3}\exp\left(\frac{r}{\rg}\right)
\left(-\dot f_R^2+\dot f_T^2\right). 
\end{equation}
Using the freedom to 
rescale the parameter $v$, 
we can set
\begin{equation}
N_\mu N^\mu=-\left[
\frac{r}{4\rg^3}\exp\left(\frac{r}{\rg}\right)\right]^2. 
\end{equation}
Then, the normalized vector which is normal to $\Sigma_1$ is given by 
\begin{eqnarray}
n^\mu=\pm\left(\dot f_R~,~\dot f_T~,~0~,~0\right). 
\end{eqnarray}
We note that the following equation is satisfied:
\begin{equation}
-\dot f_R^2+\dot f_T^2=-\frac{r}{4\rg^3}\exp\left(\frac{r}{\rg}\right). 
\label{eq:vconstraint}
\end{equation}
Due to Eq.~\eqref{eq:vconstraint}, the intrinsic metric of a CMC hypersurface is written in the form
\begin{equation}
\dd\ell^2=\dd v^2+r\bigl(f_T(v),f_R(v)\bigr)^2\dd\Omega^2.
\end{equation}
This equation implies that the parameter $v$ is the proper length measured in the radial direction, and hence we may call it the proper radial coordinate. 
See Appendix A for the metric of the spacetime in the CMC coordinate system.

For later convenience, we differentiate 
Eq.~\eqref{eq:vconstraint} with respect to $R$ or
$T$ to obtain 
\begin{eqnarray}
-\dot f_R\del_R \dot f_R+\dot f_T \del_R \dot f_T
=-\frac{\del_R r}{8\rg^3}\exp\left(\frac{r}{\rg}\right)\left(1+\frac{r}{\rg}\right), 
\\
-\dot f_R\del_T \dot f_R+\dot f_T \del_T \dot f_T
=-\frac{\del_T r}{8\rg^3}\exp\left(\frac{r}{\rg}\right)\left(1+\frac{r}{\rg}\right), 
\end{eqnarray}
where $\dot f_T$ and $\dot f_R$ are 
treated as fields on the $T$-$R$ plane.

The basic equation is given by 
\begin{equation}
K=\nabla_\mu n^\mu=3H={\rm spatially~constant}. 
\end{equation}
This equation leads to
\begin{equation}
\pm\frac{1}{r}\exp\left(\frac{r}{\rg}\right)
\left\{\del_T\left[r\exp\left(-\frac{r}{\rg}\right)\dot f_R\right]
+\del_R\left[r\exp\left(-\frac{r}{\rg}\right)\dot f_T\right]\right\}=3H. 
\label{eq:maineq}
\end{equation}
Using 
$$
\ddot f_R=\dot f_R \del_R\dot f_R+\dot f_T \del_T\dot f_R,
$$
we obtain 
$$
\del_T\dot f_R=\frac{1}{\dot f_T}
\left(\ddot f_R-\dot f_R\del_R\dot f_R\right), 
$$
and hence, 
we can rewrite $\del_T \dot f_R+\del_R\dot f_T$ as 
\begin{eqnarray}
\del_T \dot f_R+\del_R\dot f_T
&=&\frac{\ddot f_R}{\dot f_T}+\frac{1}{\dot f_T}\left(-\dot f_R\del_R \dot f_R
+\dot f_T\del_R\dot f_T\right)
\cr
&=&\frac{\ddot f_R}{\dot f_T}
-\frac{\del_R r}{8\dot f_T\rg^3}\exp\left(\frac{r}{\rg}\right)
\left(1+\frac{r}{\rg}\right).  
\label{eq:dels1}
\end{eqnarray}
By a similar procedure, we obtain
$$
\del_R \dot f_T=
\frac{1}{\dot f_R}\left(\ddot f_T-\dot f_T\del_T\dot f_T\right)
$$
and hence, we have
\begin{eqnarray}
\del_T \dot f_R+\del_R\dot f_T
&=&\frac{\ddot f_T}{\dot f_R}-\frac{1}{\dot f_R}\left(\dot f_T\del_T \dot f_T
-\dot f_R\del_T\dot f_R\right)
\cr
&=&\frac{\ddot f_T}{\dot f_R}
+\frac{\del_T r}{8\dot f_R\rg^3}\exp\left(\frac{r}{\rg}\right)
\left(1+\frac{r}{\rg}\right). 
\label{eq:dels2}
\end{eqnarray}
From Eqs.~\eqref{eq:maineq} and \eqref{eq:dels1}, 
we obtain 
\begin{equation}
\ddot f_R=\dot f_T\left[\pm K-\left(\frac{1}{r}-\frac{1}{\rg}\right)
\left(\del_T r\dot f_R+\del_Rr\dot f_T\right)\right]
+\frac{\del_Rr}{8\rg^3}\exp\left(\frac{r}{\rg}\right)\left(1+\frac{r}{\rg}\right). 
\label{eq:ddfr}
\end{equation}
Similarly, from Eqs.~\eqref{eq:maineq} and \eqref{eq:dels2}, 
we obtain 
\begin{equation}
\ddot f_T=\dot f_R\left[\pm K-\left(\frac{1}{r}-\frac{1}{\rg}\right)
\left(\del_T r\dot f_R+\del_Rr\dot f_T\right)\right]
-\frac{\del_Tr}{8\rg^3}\exp\left(\frac{r}{\rg}\right)\left(1+\frac{r}{\rg}\right). 
\label{eq:ddft}
\end{equation}
We solve these two equations with a constraint equation \eqref{eq:vconstraint}. 

To solve these equations, we need explicit expressions 
for $\del_Tr$, $\del_Rr$ and $r$. 
Differentiating Eq.\eqref{eq:defr}, 
we obtain 
\begin{eqnarray}
\del_T r&=&\frac{T}{2\rg\left(-\dot f_R^2+\dot f_T^2\right)}, 
\label{eq:delT}
\\
\del_R r&=&\frac{-R}{2\rg\left(-\dot f_R^2+\dot f_T^2\right)}. 
\label{eq:delR}
\end{eqnarray}
Combining Eqs.~\eqref{eq:defr} and \eqref{eq:vconstraint}, 
we find 
\begin{equation}
r=\rg\ln\left[T^2-R^2-4\rg^2\left(-\dot f_R^2+\dot f_T^2\right)\right]. 
\end{equation}
This final expression for $r$ is much more convenient 
than Eq.\eqref{eq:defr} in our numerical integration. 

\section{CMC Slices in the SC Universe}
\label{sec:result}
\subsection{Boundary conditions}
The cosmic time $\tau$ and the areal radius $r$ 
on the sphere 
$\Sigma_0\cap\Sigma_1$ are denoted by $\tau_{\rm b}$ and $r_{\rm b}$, respectively. 
Then, 
the Hubble constant on $\Sigma_0\cap \Sigma_1$ is given by 
\begin{equation}
H=\frac{a'}{a}=\frac{2}{3\tau_{\rm b}}
\end{equation}
and, 
from Eq.~\eqref{eq:rsol2}, we have
\begin{equation}
r_{\rm b}=\frac{3}{2}\tau_\rmc\left(\frac{\tau_{\rm b}}{\tau_\rmc}\right)^{2/3}
=\frac{3}{2}\tau_\rmc\left(\frac{2}{3\tau_\rmc H}\right)^{2/3}. 
\end{equation}

As mentioned in Sec.~\ref{sec:intro}, 
we assume that the CMC hypersurface $\Sigma_2$ in the EdS region agrees with $\tau=\tau_{\rm b}$, 
and hence $\ell^\mu$ is tangent to $\Sigma_2$ in the EdS side on $\Sigma_1\cap\Sigma_2$. 
Since the tangent space is continuous at $\Sigma_1\cap\Sigma_2$, $\ell^\mu$ is also tangent to $\Sigma_2$ on 
the Sch side.  
Since we have $e^\mu_\tau\propto\left(T_{\rm b}',R_{\rm b}',0,0\right)$ in the Sch side, the relation $e^\mu_\tau\ell_\mu=0$ leads to  
$$
\ell^\mu=C\left(R_{\rm b}',T_{\rm b}',0,0\right),
$$
where 
\begin{equation}
C=\sqrt{\frac{r\exp[r/\rg]}{4\rg^{3}\left(T_{\rm b}'^2-R_{\rm b}'^2\right)}}. 
\label{eq:C}
\end{equation}
Thus, we have 
\begin{equation}
(\dot f_T, \dot f_R, 0, 0)=-C (R_{\rm b}',T_{\rm b}', 0, 0) 
\label{eq:dotsf}
\end{equation}
on $\Sigma_1\cap\Sigma_2$, where the negative sign has been assigned, 
so that the value of $v$ increases inward.

From Eqs.~\eqref{eq:Tb} and \eqref{eq:Rb}, we obtain 
\begin{align}
T_{\rm b}'(\tau)&=\frac{1}{3\tau_\rmc}\left(\frac{\tau}{\tau_\rmc}\right)^{1/3}
\exp\left[\frac{1}{2}\left(\frac{\tau}{\tau_\rmc}\right)^{2/3}
\right]
\cosh\left[\left(\frac{\tau}{\tau_\rmc}\right)^{1/3}+\frac{\tau}{3\tau_\rmc}\right], \\
R_{\rm b}'(\tau)&=\frac{1}{3\tau_\rmc}\left(\frac{\tau}{\tau_\rmc}\right)^{1/3}
\exp\left[\frac{1}{2}\left(\frac{\tau}{\tau_\rmc}\right)^{2/3}
\right]
\sinh\left[\left(\frac{\tau}{\tau_\rmc}\right)^{1/3}+\frac{\tau}{3\tau_\rmc}\right]. 
\end{align}
By using these equations, Eq.~\eqref{eq:dotsf} gives $\dot{f}_T$ and $\dot{f}_R$ on $\Sigma_1\cap\Sigma_2$ as 
$\dot f_T=-CR'_{\rm b}(\tau_{\rm b})$ and $\dot f_R=-CT'_{\rm b}(\tau_{\rm b})$, 
while the values of $f_T$ and $f_R$ on $\Sigma_1\cap\Sigma_2$ are given as $f_T=T_{\rm b}(\tau_{\rm b})$ and 
$f_R=R_{\rm b}(\tau_{\rm b})$ with Eqs.~\eqref{eq:Tb} and \eqref{eq:Rb}. 
Then, we can integrate Eqs.~\eqref{eq:ddfr} and \eqref{eq:ddft}  
to find a CMC slice. 

\subsection{CMC slices in the Kruskal diagram}
\begin{figure}[htbp]
\begin{center}
\includegraphics[scale=1.7]{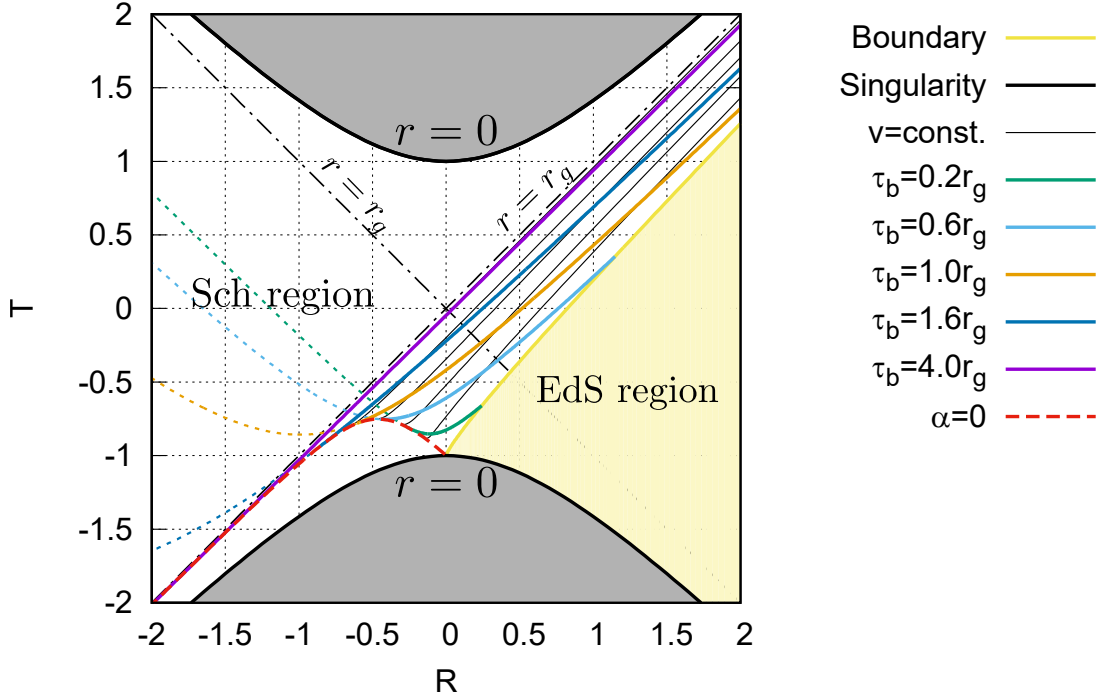}
\caption{CMC slices in the Sch region 
with the Kruskal coordinate. 
Each CMC slice is described by 
the union of a solid line segment and a dotted line segment separated by 
the $\alpha=0$ point on it: $\alpha$ is positive on the solid segment, whereas $\alpha$ is negative on the dotted segment.  
}
\label{fig:Kruskal}
\end{center}
\end{figure}

Performing numerical integrations, we finally obtain the results shown in Fig.~\ref{fig:Kruskal}. 
It is found from this figure that CMC slices do not pass through the black hole region but the white hole region. 
A similar slice is observed in an analysis of 
massless scalar field collapse in an expanding background\cite{Yoo:2018pda}. 
In order to understand the spacetime structure 
of a numerical solution, one useful way is to find marginally trapped surfaces 
associate with the outgoing or ingoing null vector fields. 
To find the marginally trapped surfaces on each spacelike hyper-surface, 
a well studied example similar to the numerical solution often plays a crucial role. Actually, 
in Ref.~\cite{Yoo:2018pda}, 
one of the present authors and his collaborators have revealed the spacetime structure by virtue of  
the results given in this paper. 
This is an example showing that 
the knowledge about the sequence of CMC slices obtained in this paper 
helps in understanding the spacetime structure of the numerical spacetime solution.

We note that, if we impose the reflection boundary condition at the center of the wormhole bridge 
as is in Refs.~\cite{Estabrook:1973ue,Nakao:1990gw,Beig:2005ef,Nakao:2009dc} differently from our present case, 
a CMC slice can pass through the black hole region. 
We also note that the sign of $K$ is fixed by the boundary condition: the slice is smoothly connected 
to the homogeneous slice in EdS where the value of $K$ is given by $K=-3H<0$. 
If we invert the time evolution, that is, considering the collapsing SC model, 
the slices do not pass through the white hole region but the black hole region(see Fig.~\ref{fig:Kruskal} flipped upside down).

Fig.~\ref{fig:Kruskal} shows that the CMC hypersurfaces 
intersect with each other. 
This fact implies that the lapse function $\alpha$ associated with the foliation by 
CMC hypersurfaces has zero points(see Appendix~\ref{appA} for details). 
In Appendix~\ref{appA}, we derive the following 
necessary and sufficient condition for the appearance of a zero point of $\alpha$:
\begin{equation}
\mathcal C:=\dot{f}_R\partial_\tau T-\dot{f}_T\partial_\tau R=0, 
\end{equation} 
where the coordinate system has been set as $(\tau,v,\theta,\phi)$ with $\tau=2K$. 
The $\tau$ derivatives $\partial_\tau T$ and $\partial_\tau R$ can be numerically calculated by 
getting a nearby CMC hypersurface specified by a slightly different value of $\tau$.  
From the results, we plot the curve $T=C(R)$ on which $\alpha$ vanishes in Fig.~\ref{fig:Kruskal}. 
We also plot the 
curves on which $v$ is constant.
In Fig.~\ref{fig:Kruskal}, each CMC slice is described by a union of solid segment and dotted segment separated by 
the $\alpha=0$ point on it. 
It can be found that 
there is no intersection between the solid segments and also no intersection between the dotted segments. 

The value of $\alpha$ can be expressed as a function of $\tau$ and $v$ as $\alpha(\tau,v)$. 
Let $v_C(\tau)$ denote the value of $v$ at which $\alpha$ vanishes, that is, $\alpha(\tau,v_C(\tau))=0$. 
The curve specified by $T=C(R)$ and $v=v_C(\tau)$ are equivalent to each other. 
Then, CMC time slicing is future directed in the domain $v<v_C(\tau)$ which is sliced by the solid segments in Fig.~\ref{fig:Kruskal}, whereas 
it is past directed in the domain $v>v_C(\tau)$ sliced by the dotted segments. 
The CMC coordinate system $(\tau, v, \theta, \phi)$  covers only the domain of 
$v<v_C(\tau)$ and $\tau>0$, or equivalently  $T<R$ and $T>C(R)$, i.e., 
the outside the black hole. 
The black hole region cannot be described by the foliation with the CMC slices 
which are smoothly connected to the homogeneous slices in EdS region. 

\section*{Acknowledgements}
This work was supported by JSPS KAKENHI Grant
Number JP16K17688 (C.Y.).

\appendix

\section{Zero points of the lapse function}
\label{appA}

A CMC hypersurface is specified by the trace of the extrinsic curvature $K=2/\tau$, or equivalently, $\tau$, 
and 
a point on the CMC hypersurface is specified by the proper radial coordinate $v$ and round coordinates, $\theta$ and $\phi$. 
We refer to the coordinate system $(\tau,v,\theta,\phi)$ as the CMC coordinate system in this paper.  
Then,  we have
\begin{align}
g_{\tau\tau}&=\left(\partial_\tau T\right)^2g_{TT}+\left(\partial_\tau R\right)^2g_{RR}=\left[-\left(\partial_\tau T\right)^2+\left(\partial_\tau R\right)^2\right]\Psi, \nonumber \\
g_{\tau v}&=\left(-\dot{T}\partial_\tau T+\dot{R}\partial_\tau R\right)\Psi, \nonumber \\
g_{vv}&=\left(-\dot{T}^2+\dot{R}^2\right)\Psi=1, \nonumber
\end{align}
where
$$
\Psi=\frac{4r_{\rm g}^3}{r}\exp\left(-\frac{r}{r_{\rm g}}\right),  
$$
a dot represents the derivative with respect to $v$ with $\tau$ fixed, 
and we have used Eq.~\eqref{eq:vconstraint} in the last equality of the third equation. 
The spatial metric $\gamma_{ij}$, its inverse $\gamma^{ij}$ and the shift vector $\beta_i$ in the CMC coordinate system are 
\begin{align}
\gamma_{ij}&={\rm diag}\left[1,r^2,r^2\sin^2\theta\right], \nonumber\\
\gamma^{ij}&={\rm diag}\left[1,\frac{1}{r^2},\frac{1}{r^2\sin^2\theta}\right], \nonumber\\
\beta_i&=\left(g_{\tau v},0,0\right), \nonumber
\end{align}
respectively, and hence we have $\beta^i\beta_i=(g_{\tau v})^2$. From $g_{\tau\tau}=-\alpha^2+\beta^i\beta_i$, we obtain
$$
\alpha^2=-g_{\tau\tau}+(g_{\tau v})^2=\Psi^2\left(\dot{R}\partial_\tau T-\dot{T}\partial_\tau R\right)^2,
$$
where we have used Eq.~\eqref{eq:vconstraint}. 
The zero point of $\alpha$ appears if and only if
\begin{equation}
\dot{R}\partial_\tau T-\dot{T}\partial_\tau R=0 \label{eq:zero-lapse}
\end{equation}
holds. This equation implies that the zero point of $\alpha$ appears if and only if the normal vector to the CMC hypersurface is orthogonal 
to the time coordinate basis $\partial/\partial \tau$.


\end{document}